\documentclass[aps,prl,preprint,superscriptaddress]{revtex4-1}

\usepackage{graphicx}
\usepackage{pdfpages}

\begin{document}


\title{Incomplete Devil's Staircase in the Magnetization Curve of SrCu$_{2}$(BO$_{3}$)$_{2}$}



\author{M.~Takigawa}
\email[]{masashi@issp.u-tokyo.ac.jp}
\affiliation{Institute for Solid State Physics, University of Tokyo, Kashiwanoha, Kashiwa, Chiba 277-8581, Japan}
\author{M.~Horvati\'{c}}
\affiliation{Laboratoire National des Champs Magn\'{e}tique Intenses, LNCMI-CNRS (UPR3228), UJF, UPS and INSA, BP 166, 38042 Grenoble Cedex 9, France}
\author{T.~Waki}
\affiliation{Department of Materials Science and Engineering, Kyoto University, Kyoto 606-8501, Japan}
\author{S.~Kr\"{a}mer}
\affiliation{Laboratoire National des Champs Magn\'{e}tique Intenses, LNCMI-CNRS (UPR3228), UJF, UPS and INSA, BP 166, 38042 Grenoble Cedex 9, France}
\author{C.~Berthier}
\affiliation{Laboratoire National des Champs Magn\'{e}tique Intenses, LNCMI-CNRS (UPR3228), UJF, UPS and INSA, BP 166, 38042 Grenoble Cedex 9, France}
\author{F.~L\'{e}vy-Bertrand}
\altaffiliation[Present address: ]{Institut N\'{e}el, CNRS and UFJ, BP 166, 38042 Grenoble Cedex 9, France}
\affiliation{Laboratoire National des Champs Magn\'{e}tique Intenses, LNCMI-CNRS (UPR3228), UJF, UPS and INSA, BP 166, 38042 Grenoble Cedex 9, France}
\author{I.~Sheikin}
\affiliation{Laboratoire National des Champs Magn\'{e}tique Intenses, LNCMI-CNRS (UPR3228), UJF, UPS and INSA, BP 166, 38042 Grenoble Cedex 9, France}
\author{H.~Kageyama}
\affiliation{Department of Energy and Hydrocarbon Chemistry, Kyoto University, Kyoto 615-8510, Japan}
\author{Y.~Ueda}
\affiliation{Institute for Solid State Physics, University of Tokyo, Kashiwanoha, Kashiwa, Chiba 277-8581, Japan}
\author{F.~Mila}
\affiliation{Institute of Theoretical Physics, \'{E}cole Polytechnique F\'{e}d\'{e}rale de Lausanne, CH-1015 Laussane, Switzerland}


\date{\today}

\begin{abstract}
We report on NMR and torque measurements on the frustrated quasi-two-dimensional spin-dimer system SrCu$_{2}$(BO$_{3}$)$_{2}$ in magnetic fields up to 34~T that reveal a sequence of magnetization plateaus at 1/8, 2/15, 1/6, and 1/4 of the saturation and two incommensurate phases below and above the 1/6 plateau. The magnetic structures determined by NMR involve a stripe order of triplets in all plateaus, suggesting that the incommensurate phases originate from proliferation of domain walls. We propose that the magnetization process of SrCu$_{2}$(BO$_{3}$)$_{2}$ is best described as an incomplete devil's staircase.
\end{abstract}


\maketitle



Coupled-dimer spin systems with singlet ground states have been extensively studied because of a variety of magnetic-field-induced quantum phase transitions. Magnetic fields strong enough to close the zero-field energy gap for the triplet excitations may simultaneously induce a longitudinal magnetization and a transverse antiferromagnetic order, breaking the $U(1)$ rotational symmetry around the field direction. This corresponds to a Bose-Einstein condensation (BEC) of triplets \cite{Giamarchi}. In the presence of frustration the magnetization curve in higher fields may show plateaus as a consequence of Wigner crystallization of triplet bosons \cite{Takigawa_Mila}.

In studying the magnetization plateaus, the layered compound SrCu$_{2}$(BO$_{3}$)$_{2}$ has played a prominent role since the discovery of a sequence of plateaus at 1/8, 1/4 and 1/3 of the saturation magnetization \cite{Onizuka001}. Its crystal structure consists of two-dimensional layers of orthogonal spin-1/2 dimers in a geometry known as the Shastry-Sutherland lattice \cite{Shastry,Kageyama}, in which frustrated interactions drastically reduce the kinetic energy of triplets, leading to the theoretical predictions of plateaus having a superstructure of localized triplets \cite{Miyahara001,Momoi001}. A symmetry-breaking superstructure of triplets has indeed been observed for SrCu$_{2}$(BO$_{3}$)$_{2}$ by copper and boron (B) NMR at the 1/8 plateau \cite{Kodama022,Takigawa041}.

More recent experiments on SrCu$_{2}$(BO$_{3}$)$_{2}$ have revealed a rich phase diagram below the 1/4 plateau, but with controversial results. B-NMR \cite{Takigawa081} and torque \cite{Levy081} measurements have revealed at least two additional phases between the 1/8 and 1/4 plateaus and excluded any plateau with a smaller fraction than 1/8. Other torque measurements, however, have been interpreted as evidence of a sequence of plateaus at 1/9, 1/8, 1/7, 1/6, 1/5, 2/9 and 1/4 \cite{Sebastian081}.
State-of-the-art analytic approaches \cite{Dorier081,Abendshein081} to determine the effective low-energy triplet Hamiltonian have confirmed the early findings of plateaus at 1/3 and 1/2 \cite{Miyahara001,Momoi001}. However, definitive conclusions regarding the plateaus and their spin structures at smaller fractions could not be reached yet because they require a precise determination of the triplet-triplet interaction at large distances, which is a very difficult yet steadily progressing task \cite{Nemec}.

In this Letter, we report on torque and $^{11}$B-NMR measurements in static high magnetic fields up to 34~T generated by a 20-MW resistive magnet at LNCMI Grenoble, and present definitive conclusions about the sequence of plateaus and their spin structures. We observed plateaus at 1/8, 2/15, 1/6, and 1/4 of the saturation and two incommensurate phases below and above the 1/6 plateau. For all plateaus, the  magnetic structures  determined by high-field NMR involve a stripe order of triplets with different commensurability, suggesting that the incommensurate phases originate from proliferation of domain walls. These findings establish the first observation of an incomplete devil's staircase in quantum antiferromagnets.

\begin{figure}[t]
\includegraphics[width=1\linewidth]{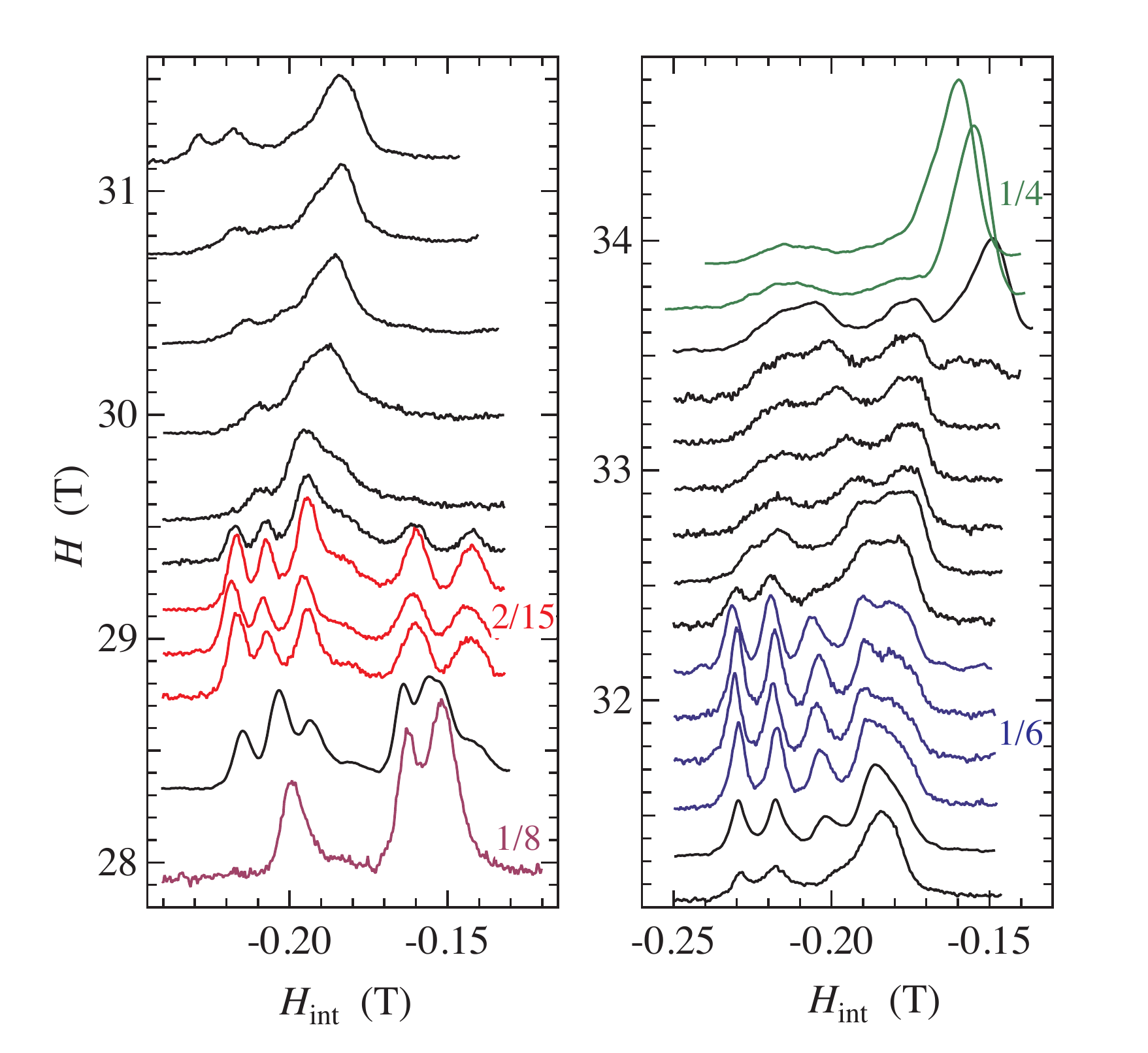}
\caption{\label{SpecAll} (color online) A part of the $^{11}$B-NMR spectra covering the most negative range of internal field obtained at $T=0.43$~K in various magnetic fields. The values of magnetic fields correspond to the position of the spectral base line on the vertical axis. The purple (below 28.2~T), red ($28.7-29.2$~T), blue ($31.5-32.2$~T), and green (above 33.6~T) spectra belong to the 1/8, 2/15, 1/6 and 1/4 plateaus, respectively.}
\end{figure}

To understand the $^{11}$B-NMR spectra, we recall that each boron site generates three NMR lines at the frequencies $f_k = \gamma \left( H + H_{\rm int} \right) + k\nu_Q$, ($k = -1, 0, {\rm or}\ 1$), where $\nu_Q = 1.25$~MHz is the quadrupole splitting for the external magnetic field $H$ applied along the $c$ axis \cite{Kodama021} and $\gamma = 13.66$~MHz/T is the nuclear gyromagnetic ratio. The internal field $H_{\rm int}$ is produced by nearby Cu spins,
\begin{equation}
H_{\rm int} = \sum_{i} A_{i} \langle S_{c}^{i} \rangle ,
\end{equation}
where $A_i$ is the hyperfine coupling constant to the $i$th Cu site that has the local magnetization $\langle S_{c}^{i} \rangle$ along the $c$ axis. We show in Fig.~\ref{SpecAll} a part of the low-frequency satellite spectra ($k = -1$) at $T = 0.43$~K covering the most negative range of internal field ($H_{\rm int} \le -0.13$~T) attributed to those boron sites whose nearest neighbor Cu carries a large magnetization. For this range of $H_{\rm int}$, the low-frequency satellite spectra do not overlap with the central line ($k = 0$) or high-frequency satellite ($k = 1$), and therefore they directly give the distribution of $H_{\rm int}$ with typical accuracy of 2~mT.

The spectrum at 27.9~T (purple) belongs to the 1/8 plateau and exhibits sharp peaks that do not move in the entire field range of the plateau, features that are characteristic of a commensurate superstructure. The peak positions agree with previous reports \cite{Kodama022,Takigawa041,Takigawa081}. Among the spectra displayed in Fig.~\ref{SpecAll}, we clearly identify two other ranges of field, $28.7-29.2$~T (red) and $31.5-32.2$~T (blue), in which the spectra present the same features, suggesting the existence of two additional plateau phases between the 1/8 and 1/4 plateaus. Outside these field ranges, the peaks are rather broad and their positions change continuously with the external field.

\begin{figure}[t]
\includegraphics[width=1\linewidth]{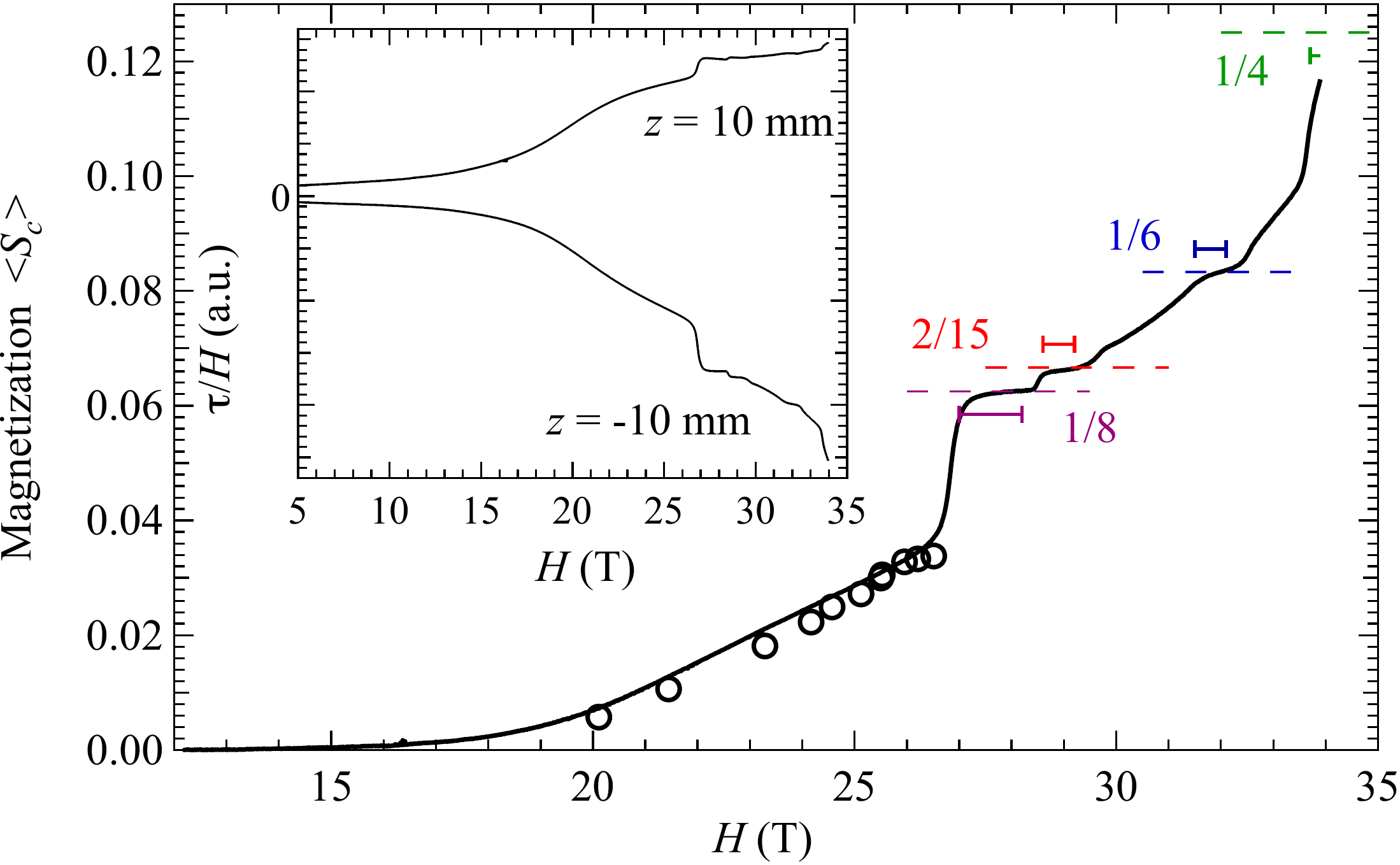}
\caption{\label{Torque} (color online) Inset: The magnetic-field dependence of the torque divided by field at $T=60$~mK for the sample positioned at $\pm 10$~mm off the nominal field center. Main panel: The thick black line represents the longitudinal magnetization (see the text) with the vertical scale appropriately adjusted. The magnetization values at 1/8, 2/15, 1/6, and 1/4 of the saturation are shown by the dashed lines. The horizontal bars indicate the field range of the plateaus determined by NMR. The open circles show the magnetization determined from the Cu-NMR shift data from Ref.~\onlinecite{Kodama022}.}
\end{figure}
The existence of new plateaus is also supported by the magnetization curve shown in Fig.~\ref{Torque} obtained by torque measurement using a cantilever technique. The noncoplanar structure of the CuBO$_3$ layers allows an intradimer Dzyaloshinskii-Moriya interaction, which has been shown to produce a transverse magnetization perpendicular to the magnetic field \cite{Miyahara071}. The torque ($\tau$) acting on the cantilever then consists of two terms: $\tau = a \mathbf{M} \times \mathbf{H} + b \left( \mathbf{M} \cdot \mathbf{\nabla} \right) \mathbf{H}$, the first one proportional to the transverse magnetization and the second one to the longitudinal magnetization \cite{Levy081}. Their relative size depends on the precise location and orientation of the sample and on the field profile inside the magnet, which are difficult to know. In the gapped phase of SrCu$_{2}$(BO$_{3}$)$_{2}$ below 15~T, where the longitudinal magnetization is strictly zero, $\tau/H$ shown in the inset of Fig.~\ref{Torque} varies linearly with $H$ due to the transverse magnetization. To eliminate this contribution and isolate that of the longitudinal magnetization, we took a linear combination of two measurements of $\tau/H$ taken at different sample positions shown in the inset of Fig.~\ref{Torque}, choosing their relative coefficients so that the resulting curve stays zero below 15~T. The result is shown as a black line in the main panel of Fig.~\ref{Torque}. It agrees very well with the magnetization determined from the Cu-NMR shift data below 26~T reported in Ref.~\onlinecite{Kodama022} (open circles).

\begin{figure*}[t]
\includegraphics[width=1\linewidth]{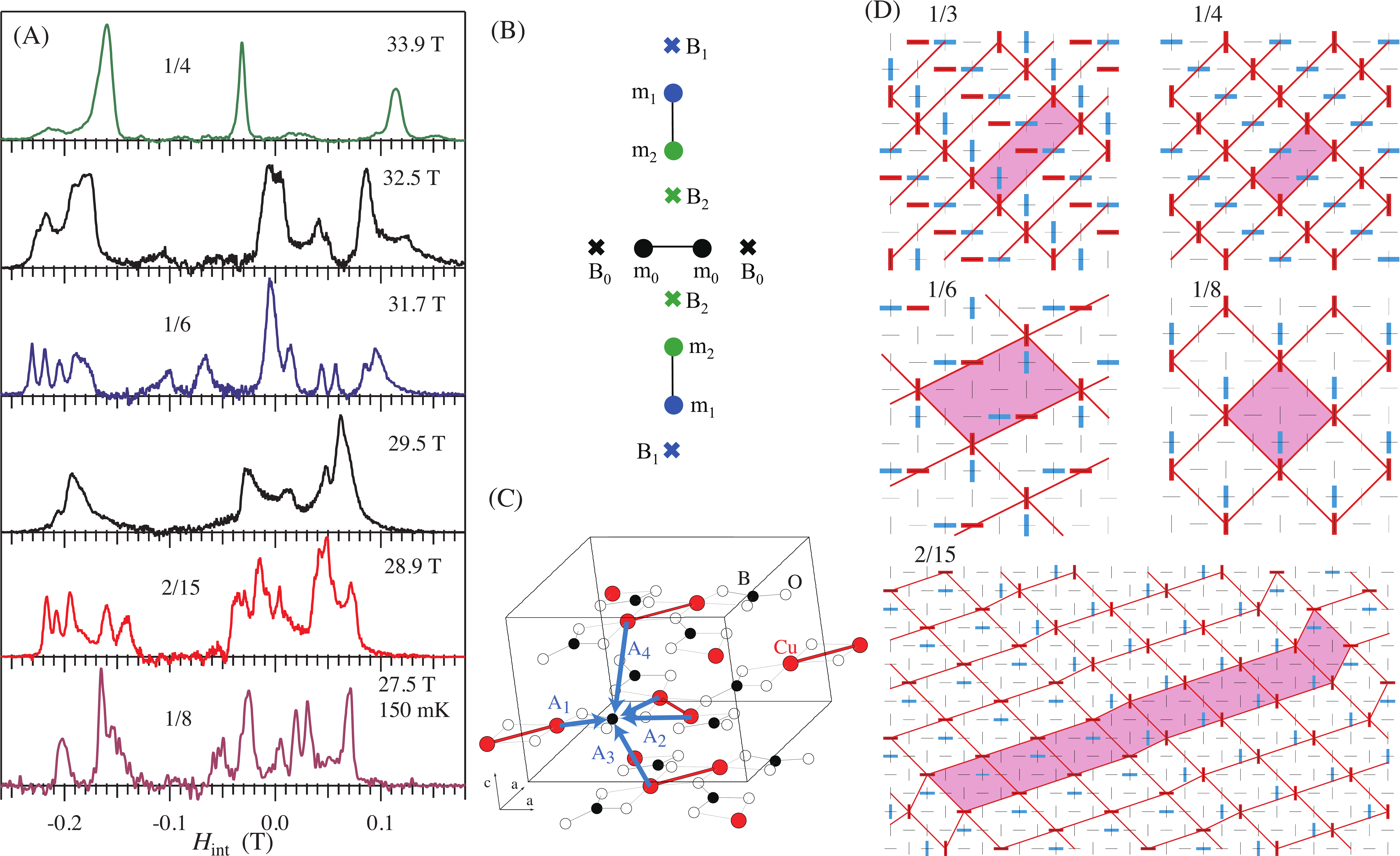}
\caption{\label{Plateaus} (color online) (A) Deconvoluted $^{11}$B-NMR spectra representing the distribution of $H_{\rm int}$ obtained at $T = 430$~mK unless explicitly indicated. The data for the 1/8 plateau at 27.5~T were taken from Ref.~\cite{Takigawa041}. (B) An extended triplet and the nearest neighbor B sites. (C) The crystal structure of SrCu$_{2}$(BO$_{3}$)$_{2}$. The arrows indicate the hyperfine coupling of a B site to the nearest and the second nearest Cu on the same layer (A1 and A2) as well as the coupling to the third and fourth neighbors on the adjacent layers (A3 and A4). (D) The spin superstructure for the plateau phases. The thin black lines show the lattice of orthogonal Cu dimers in one layer. The thick red lines show the triplet dimers carrying the largest magnetization $m_0$ in the same layer while the blue lines indicate these triplets on the neighboring layers. The unit cell of each superstructure is shown by the shaded area.}
\end{figure*}

The magnetization curve shows a series of plateaus. In addition to the 1/8 plateau and the approach to the 1/4 plateau just outside the available field range, two other plateaus can be clearly recognized: one is adjacent to the 1/8 plateau and the other is approximately halfway up to the 1/4 plateau. These field ranges agree perfectly with what we proposed from the field variation of the NMR spectra (the horizontal bars in Fig.~\ref{Torque}). The magnetizations of the first three plateaus scale as 1/8 : 2/15 : 1/6, which is partially consistent with the theoretical predictions \cite{Dorier081,Abendshein081,Nemec}. This plateau sequence is not the same as the one proposed in Ref.~\onlinecite{Sebastian081}. However, the two torque curves (Fig.~\ref{Torque} of this Letter and Fig.~1A of Ref.~\onlinecite{Sebastian081}) show anomalies at nearly identical field values, indicating that the discrepancy is not due to sample problems but due to differences in data precessing and interpretation (see the Supplemental Material A \cite{SMA}). Note that a symmetry-breaking plateau at 1/9 has already been ruled out by the previous NMR experiments \cite{Kodama022}.

Having established the sequence of plateaus, we now discuss the spin structure. The distribution of $H_{\rm int}$ was obtained by an iterative method to deconvolute the quadrupole structure from the NMR spectra \cite{Mila_unpub}. The resulting spectra are displayed in Fig.~\ref{Plateaus}(a) for all plateaus and two intermediate phases above and below the 1/6 plateau.  Let us recall that the spin density of one triplet is expected to be distributed primarily over three dimers as shown in Fig.~\ref{Plateaus}(B) \cite{Kodama022,Miyahara032}. The central dimer with a large magnetization $\langle S_c \rangle = m_0 = 0.3 - 0.4$ is surrounded by two orthogonal dimers with a negative magnetization $m_2 = -0.1$ to -0.2 (antiparallel to the external field) next to the central dimer and a positive magnetization $m_1 = 0.2 - 0.3$ at the ends. We call this cluster an "extended triplet".
It is stabilized by the cooperative (nonfrustrated) action of the polarized spins of the central triplet on the two neighboring dimers, and by its strongly reduced coupling to surrounding spins due to frustration. Other dimers have much smaller magnetization. We define B$_0$, B$_1$, and  B$_2$ as the B sites nearest to $m_0$, $m_1$, and $m_2$, respectively. All plateau spectra show a group of lines near $H_{\rm int} = -0.2$~T isolated from other lines (the part shown in Fig.~\ref{SpecAll}), which comes from B$_0$ and B$_1$ sites. 

\begin{figure}[t]
\includegraphics[width=1\linewidth]{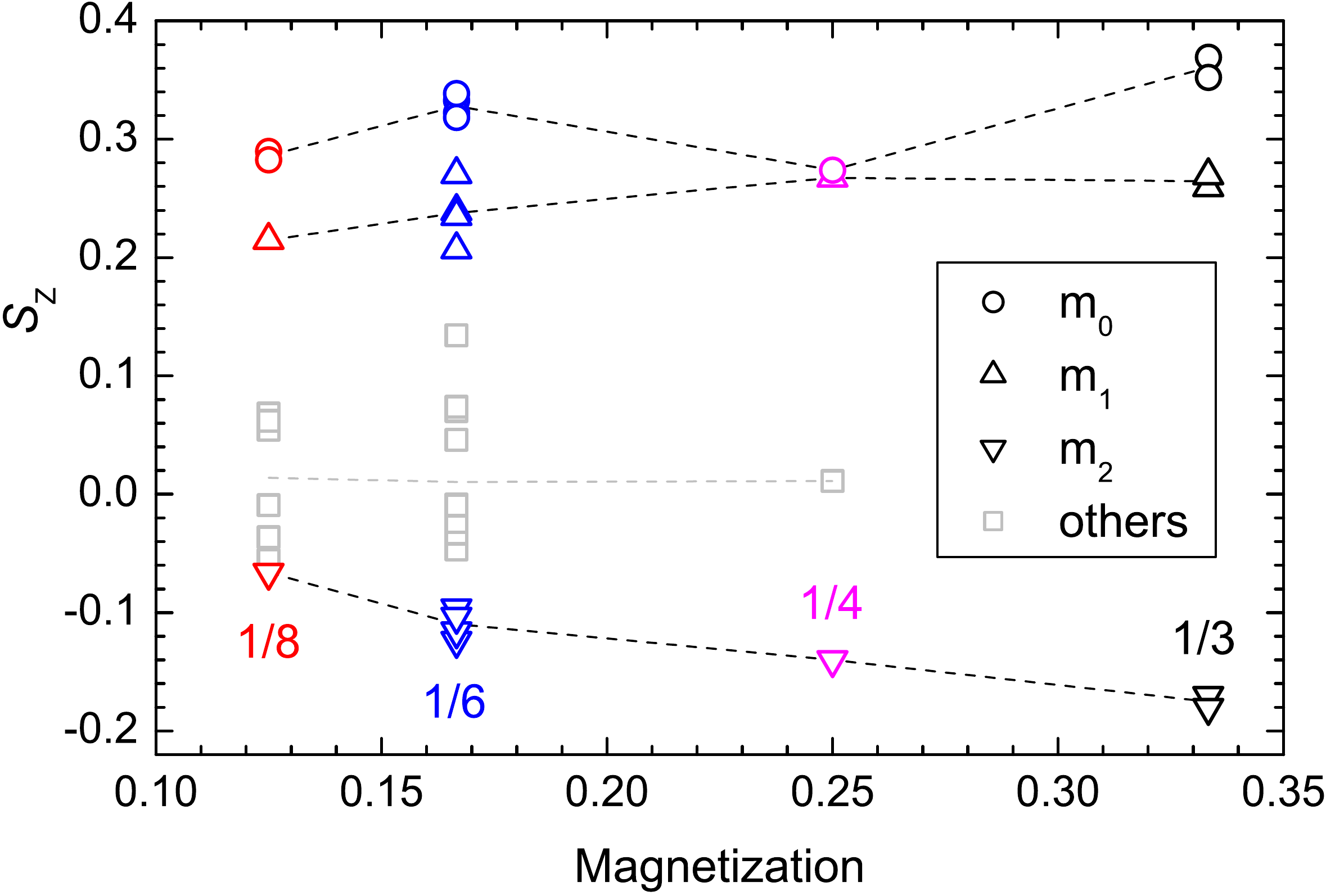}
\caption{\label{Sz} (color online) Distribution of local magnetization in the 1/8, 1/6, 1/4 and 1/3 plateau phases deduced from the peak positions of the B-NMR spectra.}
\end{figure}

The magnetic structures of plateaus were determined as follows:
(1) The coupling constant $A_i$ is the sum of the short-ranged transferred hyperfine coupling and the classical dipolar coupling. The coupling to the nearest neighbor Cu spin [$A_1$ in Fig.~\ref{Plateaus}(c)] is by far the largest. The second neighbor gives a very small contribution ($A_2$) due to the cancellation between the transferred and dipolar terms, and it is much smaller than the dipolar coupling to the third and fourth neighbors. Therefore, to zeroth order, $H_{\rm int}$ is simply proportional to the magnetization on the nearest neighbor Cu site.
(2) The zeroth-order NMR lines then shift and may be split by the dipolar field from the third and fourth neighbors, which are on the adjacent layers as shown in Fig.~\ref{Plateaus}(c). This means that the spectral features, which distinguish one plateau from the others, are primarily due to the interlayer stacking of extended triplets rather than the in-plane spin configuration. We found that the partial spectrum of B$_0$ and B$_1$ sites is specific enough in all cases to select a unique in-plane structure allowing at least one stacking pattern qualitatively consistent with the observed spectrum. Figure~\ref{Plateaus}(d) shows the structures thus determined, including the one for the 1/3 plateau which we found by applying the same analysis to the spectrum reported by Stern \textit{et al}. \cite{Stern041}. The unit cell of all these structures involve two layers as in the crystal structure.
(3) As a consistency check, we have tried to account for the entire spectrum of the 1/8, 1/6, 1/4, and 1/3 plateaus assuming the structure shown in Fig.~\ref{Plateaus}(d). For these structures, we were able to assign a particular peak value of $H_{\rm int}$ to each B site almost uniquely. The values of the local magnetization $\langle S_{c}^{i} \rangle$ have then been determined by solving Eq.~(1), since there is an equal number of inequivalent Cu and B sites in a unit cell of the superlattice (see the Supplemental Material B \cite{SMB}). The resulting values of $m_0$, $m_1$, and $m_2$, displayed in Fig.~\ref{Sz} change very little from one plateau to the next, as expected.

All plateaus show stripe order of triplets as shown in Fig.~\ref{Plateaus}(d). The structures for the 1/3 plateau and the 1/4 plateau agree with the early theoretical predictions of Refs.~\onlinecite{Miyahara001,Momoi001}. The former structure is a simple close packing of extended triplets, while in the other structures they are progressively more spaced by sites having much smaller polarization. The structure of the 1/6 plateau can be obtained from that of the 1/3 plateau by removing every other triplet from each stripe. Similarly, the square unit cell of the 1/8 plateau is obtained by removing every other triplet from each stripe of the 1/4 plateau. Note that this structure of the 1/8 plateau is different from the one with a rhomboid unit cell, which was proposed based on the previous Cu-NMR experiments \cite{Kodama022} (see Supplemental Material B \cite{SMB}). Finally, the structure of the 2/15 plateau consists of alternating domains of 1/8 and 1/6 structures. More precisely, three stripes of the 1/8 rhomboid structure (which is not the square structure of the 1/8 plateau) are followed by a single stripe of the 1/6 structure. This is a good example showing how the proliferation of domain walls leads to a structure of higher order commensurability. We should remark that the structures of the 1/6 and 2/15 plateaus are not the ones predicted theoretically in Ref.~\onlinecite{Dorier081}, presumably because the theory is valid only when the ratio of inter- to intradimer coupling is less than 0.5, while it is around 0.65 for SrCu$_{2}$(BO$_{3}$)$_{2}$. In fact, according to a recent paper \cite{Nemec}, increasing $J'/J$ and including the kinetic energy induced by Dzyaloshinskii-Moriya interactions improves dramatically the agreement between theory and experiment: a 1/8 plateau with a square unit cell is stabilized, and plateaus at 2/15 and 1/6 with stripelike structures very similar to those of Fig.~\ref{Plateaus}(d) are also present, with one of the candidates for the 1/6 plateau being exactly the structure we have deduced from NMR.

Let us now discuss the spectra in the intermediate phases below and above the 1/6 plateau.  As shown in Fig.~\ref{Plateaus}(a), they share a common feature with the spectra in the plateaus; that is, the B$_0$ and B$_1$ lines at large negative values of $H_{\rm int} \sim -0.2$~T are well separated from the rest of the spectra. This indicates that the strongly polarized extended triplets remain immobile at least within the time scale of NMR measurement (around 100~$\mu$s); i.e., they are completely localized. However, unlike in the plateaus, the spectra in the intermediate phases consist of a few broad continuous lines, which cannot be decomposed into discrete narrow lines. Such a spectral feature indicates an incommensurate magnetic structure. This is quite natural since the density of triplets, i.e., the magnetization, changes continuously, and therefore generally takes irrational values. Intermediate phases could have a commensurate structure if the system separates into two commensurate phases with their volume ratio changing with field, or if a commensurate fraction of triplets forming a superstructure coexists with the rest of delocalized triplets. However, the NMR spectra in such cases should have sharp peaks similar to the spectra in the plateau phases, clearly inconsistent with the experimental observation.

As an attempt to put the results in a broader perspective, we conjecture that the sequence of phases revealed by the present measurements is the first example of an "incomplete devil's staircase" in the context of the magnetization curve of a quantum antiferromagnet. The concept was introduced in the investigation of commensurate-incommensurate transitions: a devil's staircase is an infinite sequence of commensurate phases with increasingly large commensurability built by the proliferation of domain walls between domains of different commensurability \cite{Bak821}. It becomes incomplete when the infinite sequences of high commensurability phases with small steps are replaced by incommensurate phases \cite{Aubry}.

This interpretation is supported not only by the sequence of fractional magnetization values in SrCu$_{2}$(BO$_{3}$)$_{2}$, which is typical of a devil's staircase \cite{Bak821}, but also by the stripe structure of the plateaus determined by the present NMR experiments, which naturally allows the formation of domain walls, as seems to be the case in the 2/15 plateau. It would be very interesting to check this interpretation by a direct measurement of the wave vector of the structure, which has become accessible for the required magnetic field range by neutron scattering performed in pulsed field \cite{Matsuda}. It would also be interesting to understand the mechanism behind this devil's staircase, and whether it is connected to the recent observation of devil's staircases in the context of quantum dimer models \cite{Papanikolaou,Coletta}. Finally, it remains a challenge for theory to explain why higher commensurability
plateaus (or even lower ones such as 1/7) are unstable towards the formation of incommensurate phases.

\begin{acknowledgments}
We thank Julien Dorier, Katsuaki Kodama, Kai Schmidt and Raivo Stern for valuable discussions. This work was supported partly by a Grant-in-Aid for Scientific Research Grant (No. 21340093) from JSPS, by a GCOE program from MEXT Japan, by the French ANR project NEMSICOM, by the European Commission through the EuroMagNET network (Contract No. 228043), by the Swiss National Fund, and by MaNEP. F. M. acknowledges the hospitality of ISSP.
\end{acknowledgments}


\pagebreak


\section*{Supplementary material}

# Incomplete devil's staircase in the magnetization curve of $SrCu_2(BO_3)_2$

Supplemental Material

M. Takigawa,[1] M. Horvatić,[2] T. Waki,[3] S. Krämer,[2] C. Berthier,[2] F. Lévy-Bertrand,[2] I. Sheikin,[2] H. Kageyama,[4] Y. Ueda,[1] F. Mila[5]

[1]*Institute for Solid State Physics, University of Tokyo, Kashiwa, Chiba 277-8581, Japan*
[2]*Laboratoire National des Champs Magnétiques Intenses, LNCMI - CNRS (UPR3228), UJF, UPS and INSA, BP 166, 38042 Grenoble Cedex 9, France*
[3]*Department of Materials Science and Engineering, Kyoto University, Kyoto 608-8501, Japan*
[4]*Department of Energy and Hydrocarbon Chemistry, Kyoto University, Kyoto 608-8501, Japan*
[5]*Institute of Theoretical Physics, École Polytechnique Fédérale de Lausannne, CH-1015 Lausanne, Switzerland*

A.  Comparison of the magnetization curve with the previous torque data.

In Fig. S1, the magnetization curve (red line) presented in the main panel of Fig. 2 is compared with the torque data (grey line) shown in Fig. 1A of Ref. [12] reported by Sebastian *et al*. The magnetic-field scale of the grey curve has been compressed by 1.9 % relative to the red curve in order to make obvious the precise correspondence of anomalies in the field dependence observed in both curves, whose positions are denoted by vertical dashed lines. It is clear that in both curves the *same* sequence of phase transitions have been observed. The essential difference between the two curves is in their absolute values, originating from different treatment of the raw torque data.

As described in the main text, the red curve represents a linear combination of *two* torque measurements obtained at different sample positions, for which the coefficients were chosen to eliminate the contribution from the transverse magnetization and thus obtain the longitudinal magnetization *only*. The grey curve, on the other hand, represents a torque curve taken at one sample position [12], and is thus expected to contain the unwanted contribution from the transverse magnetization mixed together with the longitudinal magnetization. Therefore, the vertical scale for the grey curve does not necessarily represent the longitudinal magnetization, and the proposed

sequence of fractions (1/9, 1/8, 1/7, 1/6, 1/5, 2/9, and 1/4) [12], referring to longitudinal magnetization only, lose its meaning. Furthermore, removal of the transverse magnetization component in the red curve makes the plateaus and the phase transitions between them much better defined. Moreover, they correspond perfectly to what is obtained from the NMR spectra. Finally, the levels of the observed three plateaus scale *precisely* in the ratio of 1/8 : 2/15 : 1/6 fractions, and these fractions lead to a consistent determinations of the spin superstructures from the NMR spectra. In particular, for each plateau spectrum the relative weight of the integrated intensity of the strongly polarized lines validates the corresponding fractional value (see the part B of this Supplemental material), confirming that the red curve corresponds truly to the longitudinal magnetization. This is independently confirmed by the $^{63,65}$Cu NMR measurement of the absolute value of spin polarization

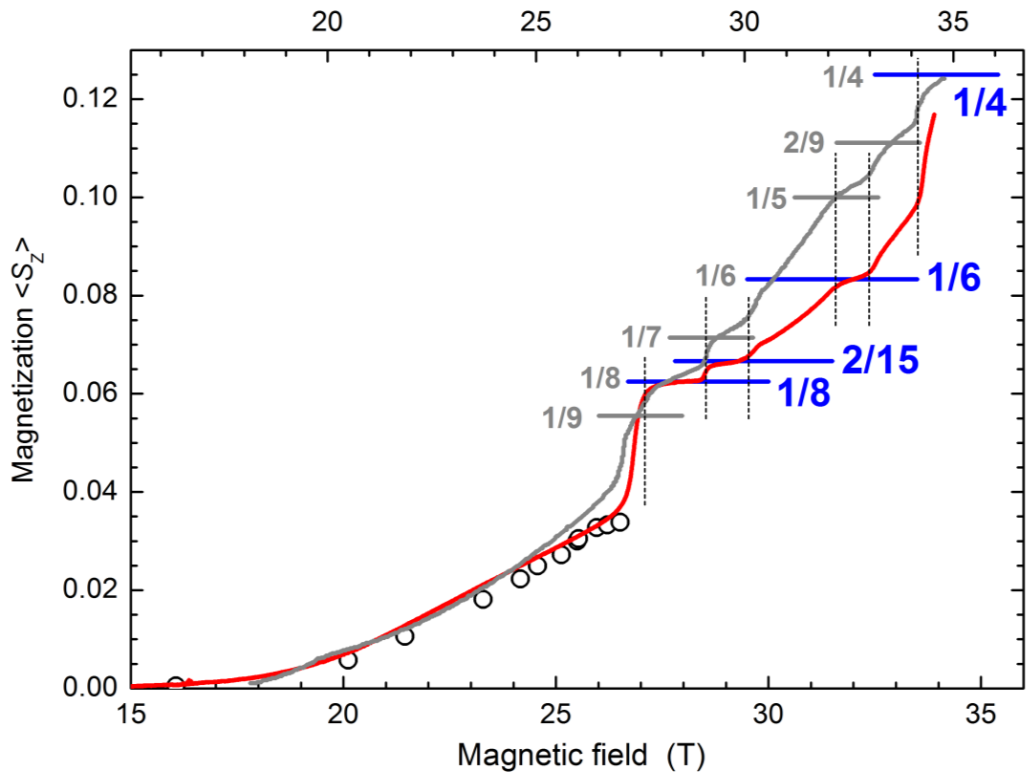

Fig. S1: The magnetization curve presented in the main panel of Fig. 2 (red line, bottom axis) is compared with the torque data shown in the Fig. 1A of Ref. [12] (grey line, top axis). The magnetization values corresponding to the plateaus proposed in this paper and in Ref. [12] are shown by the horizontal lines, while the field ranges of the plateaus are indicated by vertical dashed lines. Open circles are the $^{63,65}$Cu NMR measurement of the spin polarization [8].

up to the first plateau (open circles in Fig. S1), following closely the red curve.

As regards the difference in the field scales of the two curves presented in Fig. 1S, we remark that the magnetic field depends strongly on the position of the sample. For a typical field profile of high field magnets, having close to the field center a quadratic dependence of the amplitude 30 ppm/mm$^2$, a correction of 1.9 % corresponds to putting the sample only one inch off-center. We remark that for the red curve the calibration has been performed by metal Al NMR, and the position of the sample has been fully controlled.

Comparing in more detail the red and grey curves, we observe that the only notable qualitative difference is in the approach to the first plateau. While the red curve shows a clear sharp increase of magnetization, reflecting the mixed phase region of the first order phase transition [8], this upraise is spread over a small field interval in the grey curve and is associated to the 1/9 plateau in Ref. [12], although it is not very plateau-like. We do not really understand the origin of this anomaly; however, the NMR results [8] clearly indicate that in the corresponding field range below the 1/8 plateau phase there is only a mixed phase region with the uniform phase. There is no symmetry breaking plateau phase below 1/8 plateau, and, in particular, the existence of symmetry breaking 1/9 plateau has been ruled out. Assignment of another plateau at 2/9 in Ref. [12] merely corresponds to a minor peak in *dH/dM*. However, a peak in *dH/dM* is not a sufficient condition for a symmetry breaking plateau. Our assignment of plateaus is based not only on the magnetization curve but also on the specific features of the NMR spectra supporting a commensurate superstructure.

In summary, it is clear that the discrepancy regarding the plateau sequence between our work and Ref. [12] is not due to different properties of the samples used in both experiments but rather due to difference in the data processing and interpretation.

B. *Deduction of the distribution of local magnetization from the NMR spectra.*

Details of the analysis that has led to the determination of the spin superstructures will be the subject of a separate publication. The NMR spectra of all the plateaus, including the 1/3 plateau spectrum of ref. 17, can be shown to lead, when taken together, to a unique assessment of the spin structure of the 1/8, 1/6, 1/4 and 1/3 plateaus. Here we will present the final fits of the boron NMR spectra of these plateaus that came out of this analysis (see Figures S2 to S4). The more complicated case of the 2/15 plateau is discussed separately below (see Fig. S5).

Among various consistency checks that we have performed along the way, there is one that is particularly important, namely the possibility to account for all these spectra with similar values of

the local magnetization $m_0$, $m_1$, and $m_2$ of spins in the extended triplets shown in Fig. 3(B). The values of local magnetizations $\langle S_c^i \rangle$ at all Cu sites for the various plateaus determined from the experimental NMR spectra are shown in Fig. 4. These results were obtained as follows. First each peak in a spectrum is assigned to a particular B site. While the assignment of peaks with negative large $H_{int}$ to the $B_0$ and $B_1$ sites involves no ambiguity, assignments of other peaks to the B sites next to the weakly polarized Cu sites requires some trial and error procedure. The values of $\langle S_c^i \rangle$ can be determined by solving Eq. (1) if we know the hyperfine coupling constants $A_i$, which is the sum of the dipolar and transferred hyperfine couplings, $A_i = A_i^{dip} + A_i^{tr}$. The dipolar coupling can be calculated from the crystal parameters. The transferred hyperfine couplings are limited to the nearest and the next-nearest Cu neighbors and satisfy the condition $A_1^{tr} + 2A_2^{tr} = -0.431$ T imposed by the NMR shift data [15]. Therefore, there is only one adjustable parameter, say $A_2^{tr}$. We found that reasonable sets of values of $\langle S_c^i \rangle$ can be obtained for the range $0.05 < A_2^{tr} < 0.14$, meaning that $|A_1^{tr}| \gg |A_2^{tr}|$. The results in Fig, S5 are obtained for $A_2^{tr} = 0.09$. This plot shows that $m_0$, $m_1$ and $m_2$ evolve smoothly from one plateau to the next, $m_0$ and $m_1$ being always much more strongly polarized than the other sites, while $m_2$ gets more and more negative on going from 1/8 to 1/3 plateau, which is an interesting input for theories.

Regarding the 1/8 plateau, for which we propose a square unit cell, we note that a rhomboid unit cell has been proposed by Kodama *et al.* based on the Cu NMR spectrum [8]. However, the boron NMR spectrum of the 1/8 plateau cannot be explained by any stacking pattern of rhomboid cells. For the stacking pattern proposed in Ref. [9], for example, one of the split $B_1$ line does not match with any experimental peak. By contrast, the square cell with the stacking pattern shown in Fig. 3 and Fig. S3 fits nicely to the data, where the $B_0$ line is split by $m_0$ while $B_1$ remains unsplit. The square cell was rejected in Ref. [8] because the Cu spectrum requires more sites than allowed by the square cell (10 sites). We then reexamined the Cu spectrum and found that all peak positions of the Cu spectrum can be reproduced by the square cell with 10 sites as long as we accept pronounced disagreement on the relative peak intensity. This is definitely possible since the sensitivity of the NMR spectrometer used in that experiment may have varied significantly over the extremely wide frequency range of the Cu spectrum (100 - 400 MHz). Details will be given elsewhere.

Regarding the 2/15 plateau, a complete fit has not been attempted since only the peaks at large negative internal fields are well resolved (Fig. S5, top). Indeed, the $B_0$ and $B_1$ sites generate five sharp lines very well separated from the rest of the spectrum, which forms a continuum indicating a large number of inequivalent sites. However, the signature of the $B_0$ and $B_1$ sites is so peculiar that, after

investigating all plausible structures on the basis of our overall understanding of plateaus in this compound, we came to the conclusion that only one is qualitatively consistent with this signature. The line of argument goes as follows. First of all, the integrated intensity of the five lines at large negative internal fields is 4/15 of the entire spectrum, which is equal to the population of the $B_0$ and $B_1$ sites in the 2/15 plateau. Besides, the integrated intensity of the first set of 3 narrow lines with the most negative internal field is roughly equal to that of the second set of 2 somewhat broader lines. So we have attributed the first set to $B_0$ sites and the second one to $B_1$ sites. Now, the 3 narrow lines are split by very small internal fields. This implies that the $B_0$ sites do not have strongly polarized Cu sites as nearest neighbors in adjacent layers. By contrast, the splitting between $B_1$ lines is typical of the internal field induced by interlayer coupling to $m_2$ Cu sites. Since the spectral weights of the two $B_1$ peaks are roughly equal, half of them must be subject to such an internal field.

So, we have looked for a structure that corresponds to a magnetization 2/15 in which all $B_0$ sites are far from strongly polarized sites in adjacent layers while half the $B_1$ sites have $m_2$ Cu sites as nearest neighbors in adjacent layers. Much to our surprise, we have found a single structure that fulfills these conditions. To find this structure, we have proceeded as follows. First of all, we have looked at the structures proposed in Ref. [13], with the conclusion that none of them fulfills these conditions. Then, motivated by the stripe structure of all other plateaus, we have looked at stripe structures that correspond to a magnetization 2/15. More specifically, the 1/8 and 1/6 plateaus consist of sequences of empty and half-filled stripes, the half-filled stripes being as far as possible, a clear consequence of the repulsive triplet-triplet interaction. So we have looked for stripe structures that put triplets as far as possible while producing a magnetization equal to 2/15. All structures that fulfill these conditions consist of sequences of 15 stripes with 11 empty and 4 half-filled stripes. Among them, a single one fulfills the conditions dictated by NMR. It is represented in Fig. 3(D) and can be described as 3 stripes of the rhomboid 1/8 structure followed by 1 stripe of the 1/6 structure. In this structure, triplets in adjacent layers are parallel and sit side by side (Fig. S5, bottom), which is precisely what is needed to ensure that exactly half the $B_1$ sites have $m_2$ Cu nearest neighbors. Note that the rhomboid 1/8 structure is not the one stabilized at magnetization 1/8. Since the square and rhomboid structures have been found to have very similar energies [8, 18], the stabilization of one type of structure at the 1/8 plateau and of the other type at the 2/15 plateau is a priori possible.

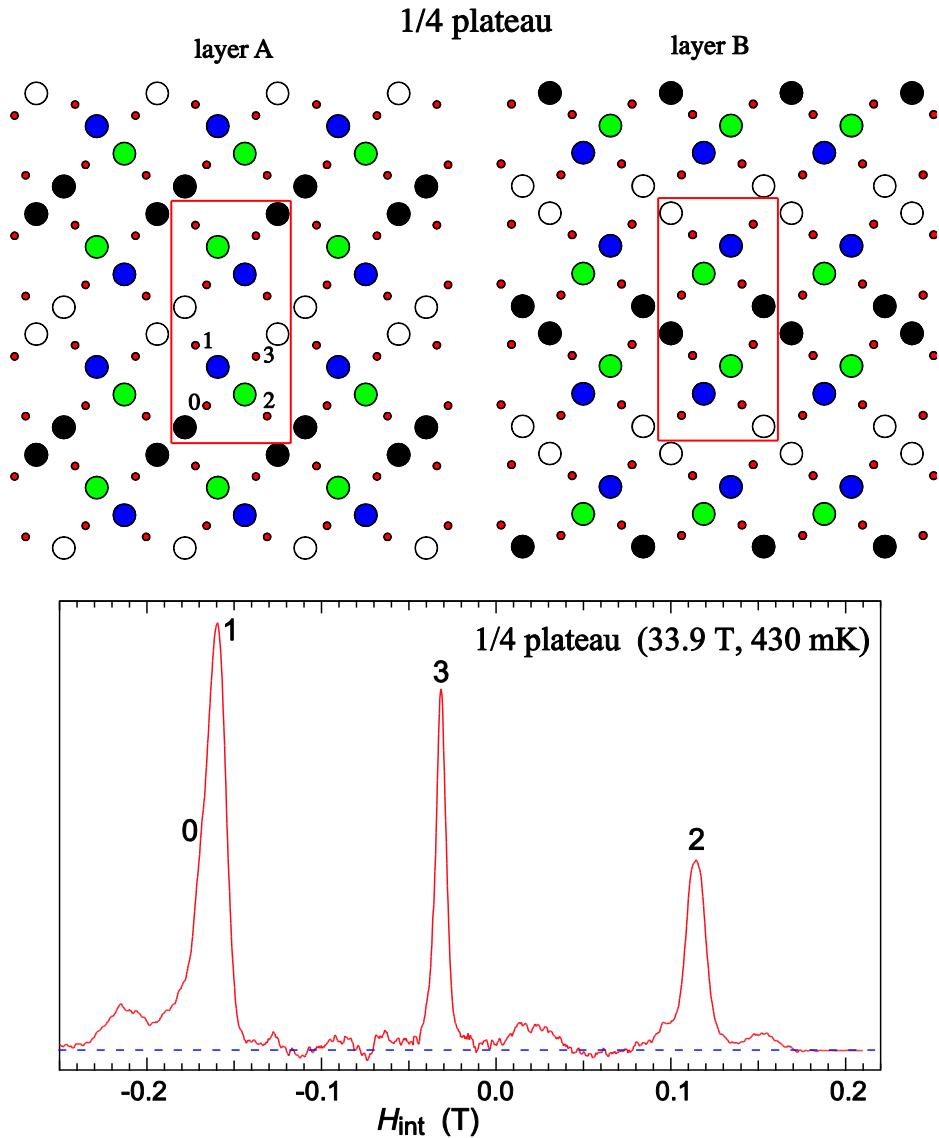

Fig. S2: Spin structure and spectrum inside the 1/4 plateau. Top: Sketch of the local magnetizations in a given layer (A) and in its two adjacent layers (B). The number of inequivalent boron sites in one layer is equal to 4, even when the interlayer hyperfine couplings are taken into account. Since all layers are equivalent up to symmetry operations of the lattice, the total number of inequivalent sites is also equal to 4. Bottom: Spectrum inside the 1/4 plateau, and line assignment. The local magnetizations obtained from the peak positions can be read off from Fig. 4. Four inequivalent sites give rise to only three peaks because the hyperfine fields are nearly equal at sites 0 ($B_0$) and 1 ($B_1$), leading to a peak with double intensity.

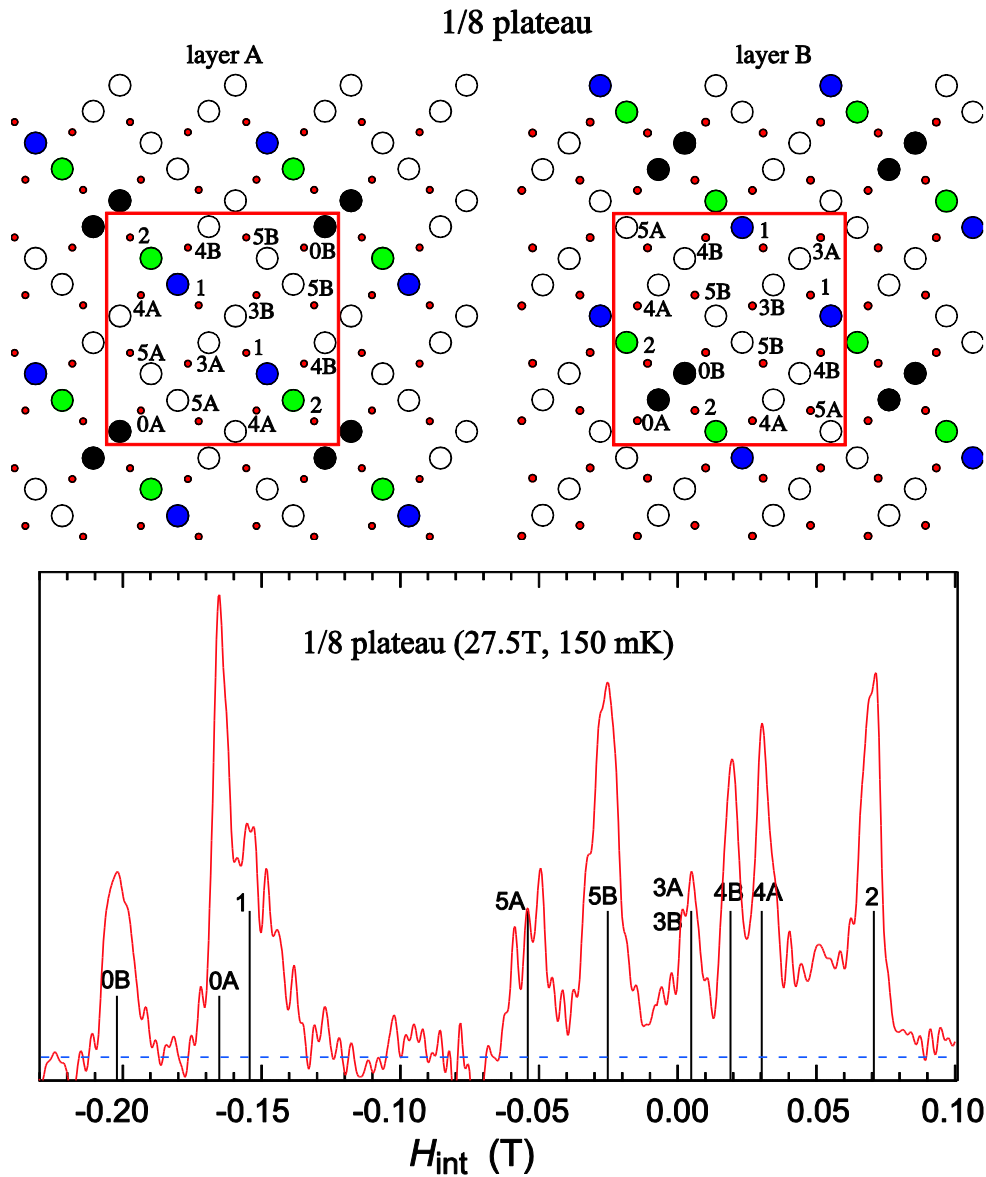

Fig. S3 Spin structure and spectrum inside the 1/8 plateau. Top: Sketch of the magnetizations in a given layer (A) and in its two adjacent layers (B) that allows one to reproduce the 9 lines of the NMR spectrum. The number of inequivalent boron sites would be equal to 6 (sites 0 to 5) if the layer were decoupled from its neighbors. When the interlayer hyperfine couplings are included, and assuming the relative positions of adjacent layers defined by (A) and (B), the number of inequivalent boron sites becomes equal to 10 (0A, 0B, 1, 2, 3A, 3B, 4A, 4B, 5A, 5B). Bottom: Spectrum inside the 1/8 plateau, and the line assignment. The local magnetizations obtained from the peak positions can be read off from Fig. 4. In our fit to the spectrum, the 10 inequivalent sites lead to only 9 lines because sites 3A and 3B have nearly equal hyperfine fields. Note that the intensities are also well reproduced, the height of the black lines being proportional to the number of sites.

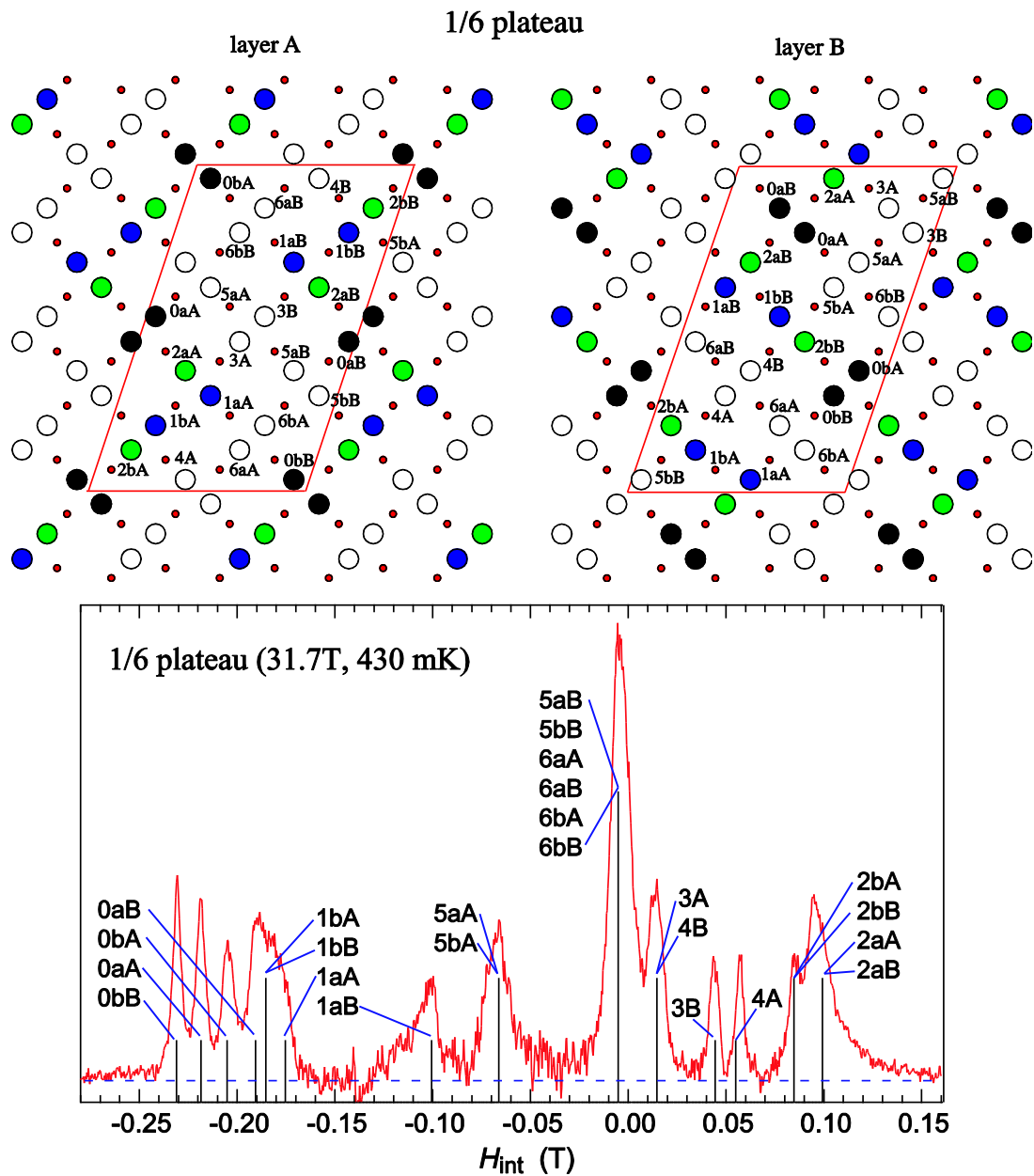

Fig. S4 Spin structure and spectrum inside the 1/6 plateau. Top: Sketch of the magnetizations in a given layer (A) and in its two adjacent layers (B) that allows one to account for the complex NMR spectrum of the bottom panel. The number of inequivalent boron sites would be equal to 12 (0a, 0b, 1a, 1b, 2a, 2b, 3, 4, 5a, 5b, 6a, 6b) if the layer was decoupled from its neighbors, where the small letter distinguishes boron sites that have the same type of nearest Cu neighbor but different environments within one layer. When the interlayer hyperfine couplings are included, and assuming the relative positions of adjacent layers defined by (A) and (B), the number of inequivalent boron sites is doubled. Bottom: Spectrum inside the 1/6 plateau, and line assignment. The local magnetizations used to fit the peak positions can be read off from Fig. 4. A careful analysis to be reported elsewhere reveals that the

presence of 4 narrow lines at the most negative end of the spectrum together with the line denoted by 1aB, which has to correspond to a 0 or 1 site by simple counting arguments although its internal field is not very negative, constitute strong enough constraints to uniquely identify the structure inside a layer.

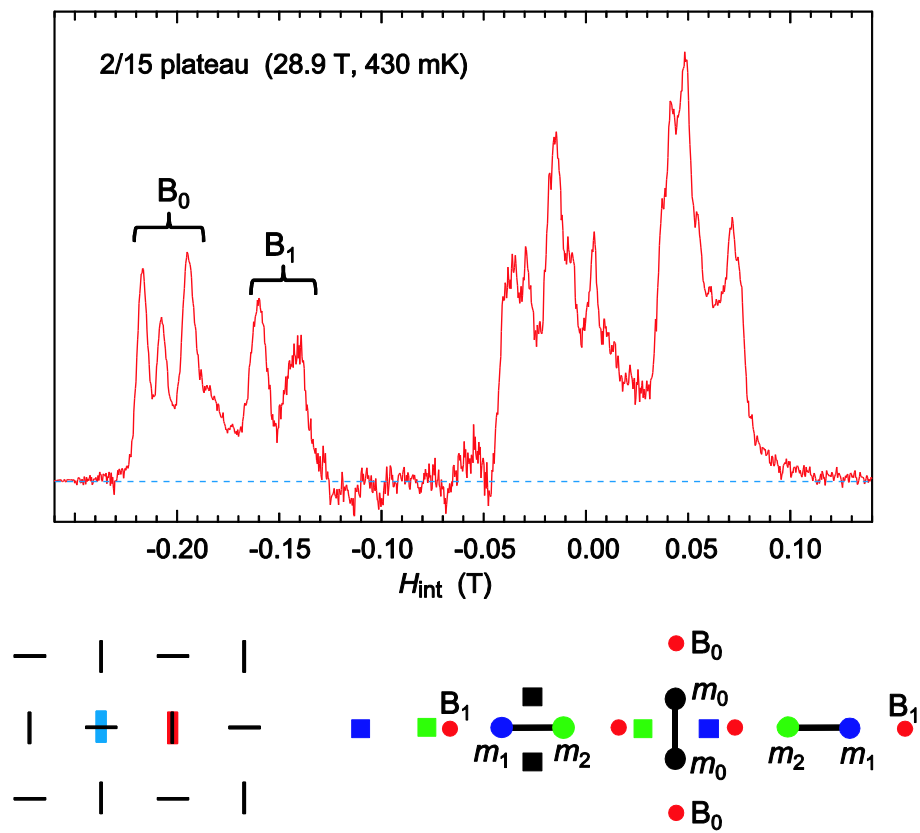

Fig. S5 Top: Spectrum inside the 2/15 plateau, and partial line assignment. Bottom left: sketch of the relative positions of strongly polarized dimers (red and blue) in adjacent layers. Bottom right: sketch of the local environment of the $B_0$ and $B_1$ sites around an extended triplet .shown by the circles. The squares show an extended triplet on the adjacent layers. The $B_1$ site on the left is close to an $m_2$ site of the adjacent layers, while the $B_1$ site on the right has only weakly polarized sites as nearest neighbors in the adjacent layers. This results into the splitting of the $B_1$ line into two lines with equal amplitude.


\end{document}